\documentclass[fleqn,usenatbib]{mnras}

\usepackage[T1]{fontenc}
\usepackage{ae,aecompl}
\usepackage{courier}

\usepackage{graphicx}	
\usepackage{amsmath}	
\usepackage{amssymb}	
\usepackage{booktabs,array,tabularx}
\usepackage{epstopdf}

\makeatletter 
\def\eqalign#1{\null\,\vcenter{\openup\jot\m@th  \ialign{\strut\hfil$\displaystyle{##}$&$\displaystyle{{}##}$\hfil      \crcr#1\crcr}}\,} 
\makeatother

\DeclareMathAlphabet{\mathsc}{OT1}{cmr}{m}{sc}
\def\testbx{bx}%
\DeclareRobustCommand{\ion}[2]{%
\relax\ifmmode
\ifx\testbx\f@series
{\mathbf{#1\,\mathsc{#2}}}\else
{\mathrm{#1\,\mathsc{#2}}}\fi
\else\textup{#1\,{\mdseries\textsc{#2}}}%
\fi}

\title{Modelling low charge ions in the solar atmosphere}

\author[R.P. Dufresne et al.]{
R.P. Dufresne,$^{1}$\thanks{E-mail: rpd21@cam.ac.uk}
G. Del Zanna$^{1}$
and P.J. Storey$^{2}$
\\
$^{1}$DAMTP, University of Cambridge, Wilberforce Road, Cambridge CB3 0WA, UK\\
$^{2}$Department of Physics and Astronomy, University College London, Gower Street, London WC1E 6BT, UK
}

\date{Accepted XXX. Received YYY; in original form ZZZ}

\pubyear{2021}

\begin{document}
\label{firstpage}
\pagerange{\pageref{firstpage}--\pageref{lastpage}}
\maketitle

\begin{abstract}
Extensions have been made recently to the coronal approximation for the purpose of modelling line emission from carbon and oxygen in the lower solar atmosphere. The same modelling is used here for other elements routinely observed in the solar transition region: N, Ne, Mg, Si and S. The modelling includes the effects of higher densities suppressing dielectronic recombination and populating long-lived, metastable levels; the presence of metastable levels typically causes effective ionisation rates to increase and recombination rates to decrease. Processes induced by the radiation field, namely photo-ionisation and photo-excitation, have been included, along with charge transfer, which occurs when electrons are exchanged during atom-ion and ion-ion collisions. The resulting ion balances are shown, and indicate significant changes compared to the frequently-employed coronal approximation. The effect on level populations within ions caused by photo-excitation is also assessed. To give an illustration of how line emission could be altered by these processes, selected line contribution functions are presented at the end.
\end{abstract}




\section{Introduction}
\label{sec:intro}

Modelling lines emitted in the solar chromosphere requires modelling which takes into consideration many factors: the principal ones being dynamical conditions in the plasma, radiative transfer, hydrogen abundance, collisions with electrons, atoms and ions, and photo-induced processes. The modelling has also been applied in certain cases to lines emitted in the transition region, such as for resonance lines in \ion{C}{ii} \citep[see, for example,][]{rathore2015,judge2003}, \ion{Si}{ii} \citep{lanzafame1994}, \ion{Si}{iv} \citep{kerr2019}, and for intercombination lines in \ion{O}{iv} \citep{olluri2013}. Such models, however, focus primarily on radiative transfer and time dependent ionisation. For the large volume of input data required for data analysis by observers, it is understandable that such detailed modelling has been typically by-passed in favour of the coronal approximation. This models emission assuming an optically thin plasma, and only considers collisions with free electrons and radiative decay in ionisation equilibrium as factors determining emission. Much focus has been placed, then, on ensuring data is as accurate and complete as possible for the coronal approximation, (see, for example, \citealt{jordan1969}, \citealt{arnaud1985} and \citealt{mazzotta1998}).

As coronal modelling is applied to lines which form deeper in the atmosphere, it is understandable that discrepancies will arise at some point. The question is where that point occurs. Once that question has been established, it raises another question: whether chromospheric-type modelling must be employed at that point, or whether there is an intermediate form of modelling which suitably describes the region. If the latter type of modelling exists, depending on its nature, it may be possible to adapt the large scale codes which currently employ the coronal approximation without the need to switch to complex, chromospheric-type modelling. 

Early models show how ion balances derived from the coronal approximation alter when the higher densities of the solar transition region are taken into consideration. The pioneering work of \cite{burgess1969} shows that the fractional ion populations are shifted to lower temperature when dielectronic recombination is suppressed. The formative work of \cite{nussbaumer1975} extends this, by showing how the long-lived, metastable levels just above the ground in carbon become populated in higher density regions, and cause the ion balance to shift to lower temperatures. They also indicate how the low charge states of carbon are affected by photo-ionisation. \cite{nussbaumer1975} go on to close their work by stating that all ions which form in the transition region would be affected by the processes they include. In another pioneering work, \cite{baliunas1980} show how the ion balance for the low charge states of silicon (up to \ion{Si}{iv}) substantially alters when charge transfer is taken into consideration.

The present work is part of a series of papers with the aim of building on those pioneering works, in order to answer the questions raised above, and to provide modelling suitable for the transition region. \cite{dufresne2019} and \cite{dufresne2020} develop the coronal approximation of carbon and oxygen by adding the effects modelled by \cite{burgess1969} and \cite{nussbaumer1975}. It is shown that, by including the effects of density on the electron collisional processes, predicted intensities are in better agreement with observations for the lines emitted below 100\,000\,K, when compared to coronal approximation modelling. However, some discrepancies still remain in the predictions for those lines. To try to resolve this, other atomic processes not normally modelled in the coronal approximation are added to the carbon and oxygen models by \cite{dufresne2020pico}. The new processes are photo-ionisation and -excitation, charge transfer and inelastic collisions with hydrogen, which all potentially become important in the cooler temperatures of the lower atmosphere and closer to the solar disc. The processes are routinely employed in chromospheric modelling. With all of the new processes added, the ion balances of both elements are considerably altered in comparison to coronal modelling for neutral to doubly-charged species, which all form below 100\,000\,K.

The aim of this work is to extend the new modelling to other elements which are routinely observed in the solar transition region. The new elements modelled in this work are: nitrogen, which lies between carbon and oxygen and could be affected in a similar way; neon, which is an element with an high first ionisation potential and can help assess whether such elements are affected by the new modelling; magnesium and silicon, both of which are low first ionisation potential elements; and, sulphur. With all of these elements modelled, the results can be used in conjunction with observations to diagnose a differential emission measure of the transition region, and then to calculate predicted intensities from the modelling. This will help assess the accuracy of the modelling, and answer the questions raised about where the coronal approximation becomes insufficient and if an intermediate form of modelling may be employed. Using the models developed in this series of papers to predict intensities and to compare them with observations, will form the basis of a separate work following this one.

In the next section of this work an outline of the new atomic processes and their reasons for being incorporated is given, as well as the methods and data used to include them. Section\;\ref{sec:ionresults} shows how the ion balances for the elements alter when the new processes are added, while Sect.\;\ref{sec:levelresults} discusses the changes to level populations and gives an assessment of how line emission may be affected as a result. Conclusions highlighting the main effects on the modelling are given at the end.

\section{Methods and data}
\label{sec:methods}

Detailed reasons and methods for including the new effects are discussed in Sect.\;2 of \cite{dufresne2020pico}. In outline, the primary reason for including photo-induced processes is that their rates are independent of temperature and density. Although density is greater in the lower solar atmosphere, the temperature is lower and the electron impact excitation and ionisation rates drop off rapidly. Wavelengths required to initiate the photo-induced processes in low charge ions are longer, and coincide with the spectrum where the solar radiance is much stronger. 

Charge transfer can occur for an element A in charge state $+z$ with hydrogen, for example, through the following reaction,

\begin{equation}
 A^{+z}~+~H~\rightarrow~A^{+(z-1)}~+~H^+~.
 \label{eqn:ct}
\end{equation}

\noindent Because of the scarcity of other elements, charge transfer through collisions with other elements can be neglected, with the exception of helium. The reverse reaction, charge transfer ionisation, may also occur. The threshold energies for charge transfer processes are smaller by the ionisation potential of H or He, depending on which is involved in the collision. Thus, the charge transfer (CT) rates are potentially much stronger than free electron rates at low temperature. \cite{dufresne2020pico} also explored the effect of inelastic collisions between neutral oxygen and neutral hydrogen, and found that it is not an important process at the temperatures considered in these models; the \ion{H}{i} abundance relative to free electrons is much diminished higher up in the chromosphere. This latter process will not, therefore, be explored further here. 

Concerning free electron processes in the lower atmosphere, \cite{dufresne2019} and \cite{dufresne2020} show for carbon and oxygen that suppression of dielectronic recombination (DR) and ionisation and recombination from metastable levels notably alter ion formation. The latter effect is relatively stronger than the former in altering the charge state distribution for the first few charge states. To model ionisation from metastable levels, for which data are not generally available compared to the other processes included, an approximation is used. This makes it simple to apply to the many ions being modelled in this work. This approximation will be discussed next, followed by a description of how the models have been built.

\subsection{Collisions involving free electrons and radiative decay}
\label{sec:lrmethods}

\subsubsection{Ionisation}

For all the elements explored in this work, (N, Ne, Mg, Si and S,) collisional-radiative (CR) models were built on the same basis as in the earlier works, apart from the electron impact ionisation (EII) data. For the ground levels of all ions the EII rate coefficients of \cite{dere2007} were imported from \textsc{Chianti}. Producing EII rate coefficients for the metastable levels of all the included ions by \textit{ab initio} methods would be an onerous task, especially because excitation--auto-ionisation would be required for many ions. Another complication to this is that readily-available codes for EII use the distorted wave approximation. This has a large degree of uncertainty for low charge ions, and the results in each case may need adjusting to compare favourably with experiment. To circumvent these challenges, an approximate method to determine EII rates for metastable levels has been used. 

\begin{figure}
 \centering
 \includegraphics[width=8.5cm]{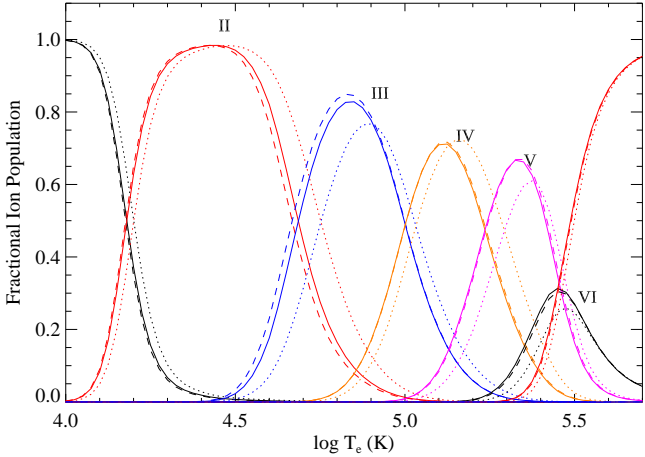}
 \caption{Electron collisional model for oxygen at $10^{12}$~cm$^{-3}$ density: solid line - Dufresne et al., dashed - using Burgess \& Chidichimo method to estimate metastable EII rate coefficients, dotted - Dufresne et al. coronal approximation.}
 \label{fig:ocrmscaling}
\end{figure}

In many works, EII rate coefficients are obtained from long-established approximations, such as those produced by \cite{lotz1968}, \cite{burgess1964ecip}, \cite{burgess1983}, \cite{vriens1980} and \cite{voronov1997}. All were developed for a certain task and have limitations. For instance, \cite{burgess1977} state that the much-used, exchange classical impact parameter (ECIP) method \citep{burgess1964ecip} was developed to produce rates for Rydberg levels at solar corona temperatures, which would not be suitable for this work. In order to obtain more accurate data for metastable states, rate coefficients were obtained in the following way. The \cite{burgess1983} approximation was used to calculate rate coefficients for the ground and metastable level based on their ionisation potentials and effective number of electrons; the ratio of the metastable to ground rate coefficients was determined; and then, the ground level rate coefficients of \cite{dere2007} were multiplied by the ratio, giving an estimated rate coefficient for the metastable level.

\cite{dere2007} included excitation--auto-ionisation (EA), and so this process is reflected in the results, although the EA contribution can vary to a different degree between levels than direct ionisation. The approximation from \cite{burgess1983} was used because it includes the Wannier factor, which better reflects ionisation at lower energies, making it more suitable for temperatures in the lower solar atmosphere. Each ion in the isoelectronic sequences of second row elements was treated as a Case (ii) ion, as defined by \cite{burgess1983}, and those of the third row isoelectronic sequences were treated as Case (i) ions, as long as they were of charge $+2$ and higher. The only unknown factor in the Burgess \& Chidichimo approximation is the quantity C, which is normally empirically determined. Because only the ratio between two levels is used, the constant cancels out. It is similar to the method used in \cite{dufresne2019} for \ion{C}{i}, and for \ion{O}{ii} and \ion{O}{iii} in \cite{dufresne2020}, except in those works the adjustment was made to experimental results. Here, the ground level cross sections have already been adjusted by \cite{dere2007} to reflect experiment. It is also, in essence, much the same as that used by \cite{giunta2011}, who uses Burgess \& Chidichimo for the metastable rate coefficients, having determined a suitable value for the quantity C by comparing the distorted wave calculation for the ground level with that from Burgess \& Chidichimo.

Figure~\ref{fig:ocrmscaling} shows the outcome of this method when it is used in the CR model for oxygen. To make a like-for-like comparison, the ratio of the metastable to ground rate coefficients from the Burgess \& Chidichimo approximation was used to multiply the total, ground level rate coefficients from \cite{dufresne2020}. These were run through the CR model of the same work, and the results shown for when the actual and estimated rates are utilised. This is shown for an electron density of $10^{12}$\,cm$^{-3}$ to ensure the metastable levels are fully populated, which highlights the greatest difference this method produces. It can be seen that this method reproduces the results for oxygen very well. The only noticeable variance from the actual ion balance is for \ion{O}{ii} ionising to \ion{O}{iii}. This is most likely because the \ion{O}{ii} ionisation rate coefficients in \cite{dufresne2020} were adjusted to experiment, and so two approximations are being compared with one another, rather than a comparison with \textit{ab initio} methods.

\subsubsection{Recombination}

The total recombination rate coefficients for ground and metastable levels were obtained from \cite{badnell2006}, \cite{kaur2018}, \cite{abdel-naby2012} and \cite{altun2007} for radiative recombination (RR), and from the following sources for DR: \cite{kaur2018}, \cite{abdel-naby2012}, \cite{altun2007}, \cite{altun2006}, \cite{zatsarinny2004}, \cite{zatsarinny2003}, \cite{zatsarinny2006}, \cite{mitnik2004}, \cite{zatsarinny2005}, \cite{altun2004}, \cite{colgan2003}, \cite{colgan2004}, \cite{bautista2007}, and \cite{badnell2006dr}. The only missing recombination rates for the metastable levels are for \ion{S}{ii} recombining into \ion{S}{i}. A new calculation was made for this case using \textsc{Autostructure} \citep{badnell2011}, and this was incorporated for the ground and metastable levels of \ion{S}{ii}. The details of this calculation will be the subject of a subsequent paper.

\begin{figure}
	\centering
	\includegraphics[width=8.4cm]{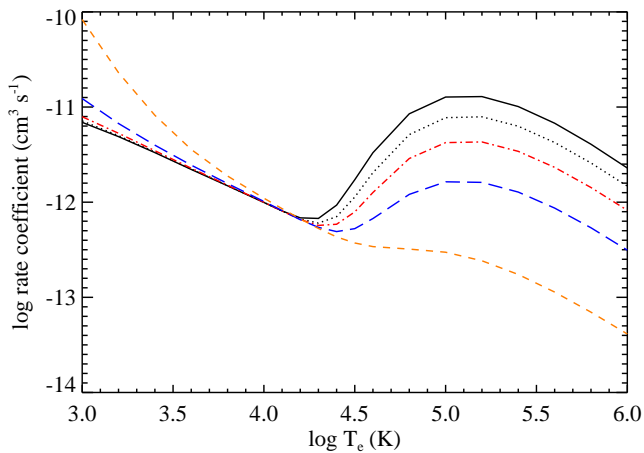}
	\caption[width=1.0\linewidth]{Effective recombination rate coefficients from a CR model with $n$-dependent Rydberg states for \ion{S}{ii}, at various densities: black solid line - $10^2$\,cm$^{-3}$ density, black dotted - $10^6$\,cm$^{-3}$, red dot-dashed - $10^8$\,cm$^{-3}$, blue long dashed - $10^{10}$\,cm$^{-3}$, orange short dashed - $10^{12}$\,cm$^{-3}$.}
	\label{fig:s2effrates}
\end{figure}

DR was suppressed for all elements in the same way as the earlier works on carbon and oxygen. This was achieved by taking, at each temperature in the model, the ratio of the \cite{summers1974} effective recombination rate coefficient at the density of interest to the Summers rate coefficient at the lowest density ($10^4$\,cm$^{-3}$). The Summers data were obtained from the `\texttt{.acd}' ADF11 format files on the OPEN-ADAS website\footnote{open.adas.ac.uk}. However, in the present calculation, the DR rate coefficients for \ion{S}{ii} are significant at low temperatures, below 10000K. This is caused by the contribution from autoionising states with relatively low principal quantum number lying very close to the ionisation threshold of \ion{S}{i}. This is not suppressed at high densities in the same way as the high temperature component. To provide an estimate of how the DR rate coefficients would be suppressed with density, the new level-resolved data were added to an $n$-dependent collisional-radiative model which includes all Rydberg levels, using the rates and techniques described by \cite{storey1995}. This model assumes that the $nl$ Rydberg states are populated according to their statistical weights, which is a good approximation for those states where DR is the dominant recombination mechanism and where collisional suppression is occurring. How the effective rate coefficients change with density is shown in Fig.\;\ref{fig:s2effrates}. The change at each temperature and density, relative to the low density case, was used to suppress the new \ion{S}{ii} DR rate coefficients in the present model. The new suppression factors at various densities are made available online.

As discussed in \cite{dufresne2020pico}, three-body recombination was tested on carbon and oxygen and was found to have no influence on the ion balances. It was also tested here on the included levels of nitrogen; the rates were far lower than for the other recombination processes. The process has most effect on Rydberg levels, which are not included in these models. The effect of three-body recombination can be seen in Fig.\;\ref{fig:s2effrates}, where the effective recombination rate coefficients for \ion{S}{ii} from the \cite{storey1995} collisional-radiative model are noticeably enhanced below $10^4$\,K at $10^{12}$\,cm$^{-3}$ density. This density is significantly higher than those relevant for the present work, but the process may be important when modelling ion formation in such locations as solar active regions and flares.

\textbf{}

\subsubsection{Bound-bound transitions}

All of the radiative decay and proton and electron excitation rates required to determine the level populations were imported from \textsc{Chianti} v.9 \citep{dere1997, dere2019}, with a few exceptions. There are a limited number of radiative decay rates in \textsc{Chianti} for \ion{N}{i}. More transitions were included from \cite{tachiev2002}, the original source of the \textsc{Chianti} data. \textsc{Chianti} does not contain any internal transition rates for \ion{Si}{i}, and so a new model was built. Level energies were drawn from NIST \citep{kramida2018}; all terms up to $3s^2\,3p\,3d\;^3D^o$ were included. The decay rates were taken from the multi-configuration Hartree-Fock (MCHF) results of \cite{fischer2005}. Further decay rates for these configurations were included from an MCHF calculation using the same methods as \cite{fischer2005} and the updated version of the MCHF codes, \textsc{ATSP2K} \citep{fischer2007atsp}. The decay rates from this calculation were within a few per cent of almost all of those from \cite{fischer2005}. 

After making enquiries, the electron impact excitation (EIE) cross sections of \cite{gedeon2012} for \ion{Si}{i} are no longer available, and so a new calculation was carried out using \textsc{Autostructure} \citep{badnell2011}, which implements the distorted wave (DW) approximation. Scaling parameters which minimise the level energies were obtained for the orbitals, and level energy corrections using the NIST values were applied. Comparing the results with the $LS$-coupling, singlet transitions presented by \cite{gedeon2012} shows that the present collision strengths were, in many cases, within 25 per cent of their R-Matrix without pseudo-states calculation. Although DW is not usually reliable for neutral states, its only purpose in this work is to establish metastable level populations so that ionisation from those levels can be properly accounted for. The models for ion balances are not as sensitive to excitation rates as those which are required for calculating line intensities.

\subsection{Photo-induced processes}
\label{sec:pimethods}

\begin{figure*}
 \centering
 \includegraphics{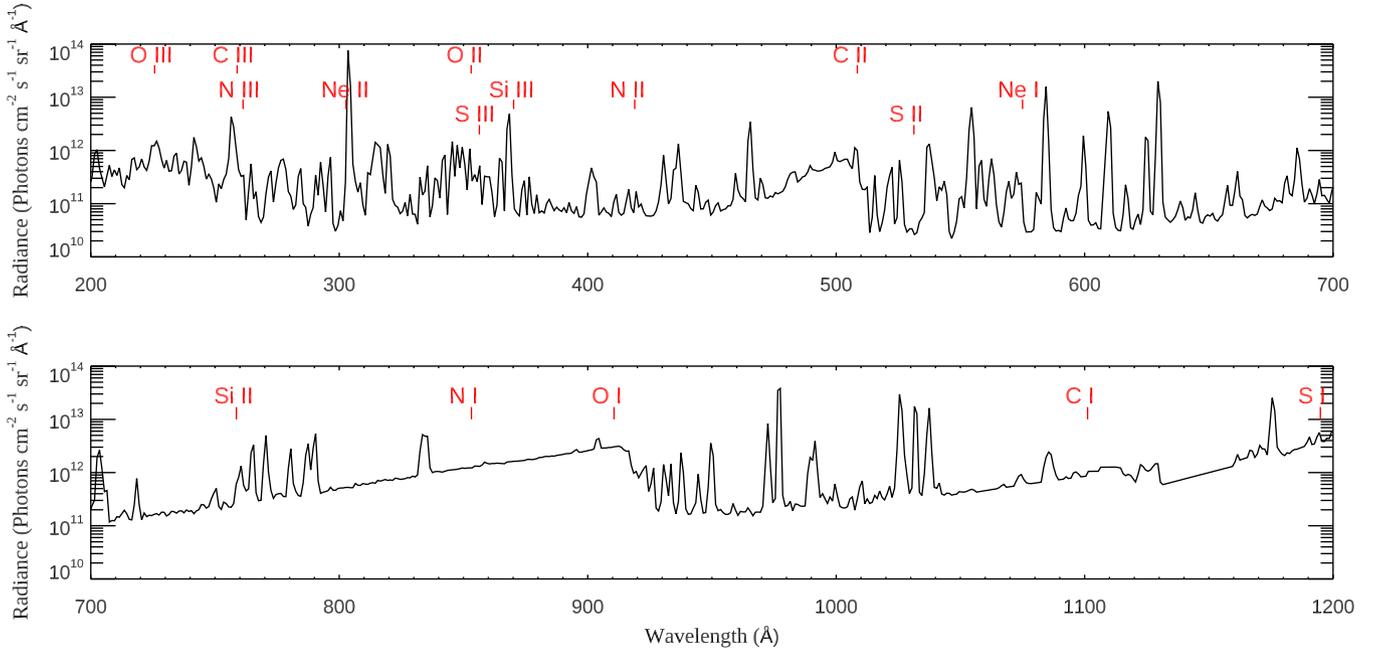}
 \caption{The radiances used for the photo-induced processes, derived from the Woods et al. Whole Heliospheric Interval irradiance spectrum. The thresholds for photo-ionisation of the low charge ions are also indicated.}
 \label{fig:solarflux}
\end{figure*}

The distorted wave PI cross sections of \cite{badnell2006} were used up to Mg-like ions, as made available on the Atomic Processes for Astrophysical Plasmas (APAP) website\footnote{www.apap-network.org}. Above that sequence, Opacity Project (OP) cross sections were used. The only available cross sections for the latter ions were in $LS$-coupling. Having checked the cross sections for $LS$-coupling and $LSJ$-coupling from \cite{badnell2006}, it was found that the $LSJ$-coupling cross section for each level within a term are identical to the $LS$-coupling cross section for the term, (within the uncertainty of the two calculations). Therefore, the OP cross section for each term was used for every level in that term in the model. The OP cross sections were adjusted in the usual manner: the energies for the cross section were shifted by the difference between the OP threshold and that from NIST, and any cross sections below the threshold were cut off. Testing the resulting PI rates for the ground level of \ion{Si}{iii} using the cross sections from \cite{badnell2006}, \cite{verner1996} and the OP gives differences of 23 per cent; for the metastable levels the maximum difference is 28 per cent. This is within the differences obtained for the rates when using the $LS$- and $LSJ$-coupling cross sections from the same work of \cite{badnell2006}. Photo-excitation and stimulated emission rates were added for all transitions for which there are radiative decay rates in \textsc{Chianti}.

The same radiances as the earlier work were used to determine rates for the photo-induced processes. These were derived from the irradiances of the Whole Heliospheric Interval reference spectrum of \cite{woods2009}, obtained from observations of the quiet Sun. Above this, for wavelengths longer than 24000\,\AA, which is relevant for photo-excitation, the spectrum was supplemented by a blackbody spectrum with a temperature of 6100\,K. A portion of this spectrum, for the region relevant for photo-ionisation of the low charge states, is shown in Fig.\;\ref{fig:solarflux}. The PI threshold wavelengths of the ions are also indicated.

\subsection{Charge transfer}
\label{sec:ctmethods}

In this work, charge transfer rate coefficients from quantum mechanical calculations were sought; these are more suitable than more simplified treatments for lower temperature scenarios, including the Sun. All of the CT rate coefficients in the sources are given in $LS$-coupling. \cite{stancil1999o1} shows that the $LS$-coupling, CT rate coefficients are obtained from the sum of the $LSJ$-coupling rate coefficients for each level weighted by the populations of the level. Therefore, for these models, for each initial level in a term, the $LS$-coupling rates were split by statistical weight because most ions affected by CT generally have their fine structure levels populated according to statistical weight in the solar atmosphere. Following the modelling in \cite{dufresne2020pico}, the ion fractional abundances of H at each temperature were taken from the model atmosphere of \cite{avrett2008}, as well as the ratio $\frac{Ab(H)}{N_e}$. For CT processes with He, the abundance was taken from \cite{asplund2009} and the ion fractional populations from \textsc{Chianti}. Since the He ion populations are from the coronal approximation, their effect on CT in the model were checked against those predicted by the new model of \cite{delzanna2020} at a few temperatures. Although there were significant differences in the He ion populations in some cases, for the ion most affected by CT with He, \ion{Si}{iv}, its fractional population was lowered by 8 per cent at the peak when using the \cite{delzanna2020} results. At other temperatures it was affected by significantly less than this, and so the \textsc{Chianti} populations were retained. The sources of the charge transfer rate coefficients used in the models are now listed.

\subsubsection{Nitrogen}

\cite{lin2005} provides CT rate coefficients for ionisation and recombination between \ion{N}{i} and \ion{N}{ii}; they go up to 200000~K, which is more than sufficient for these ions. They include rate coefficients for transitions into and out of the ground and metastable levels of \ion{N}{i}, but do not consider transitions to and from the metastable levels of \ion{N}{ii}. \cite{kimura1997} considered the metastable, $2s^2\,2p^2\;^1D$ term in \ion{N}{ii} in their calculations, but do not show the results. They state that CT from metastable terms will become strong only at higher energies. Since the region in which metastable ionisation for oxygen becomes notable is 40\,eV and the threshold for the transition in nitrogen is higher, it is, therefore, assumed that metastable transitions from \ion{N}{ii} to \ion{N}{i} will only be relevant for temperatures far above that of the solar atmosphere. \cite{barragan2006} considered CT from the ground and metastable levels of \ion{N}{iii}, and their rate coefficients are included here. CT ionisation out of \ion{N}{ii} is not relevant because CT from \ion{N}{iii} recombines into levels which decay on timescales far shorter than CT ionisation rates. For \ion{N}{iv}, the cross sections of \cite{bienstock1984} were supplemented at low energies by the results of \cite{gargaud1981} in order to give rates which are relevant at the formation temperature of \ion{N}{iii}. CT with He from \ion{N}{iv} was included using the rate coefficients of \cite{liu2011}. They also report double electron capture into \ion{N}{ii}, but these rate coefficients are several orders of magnitude lower than single electron capture, and so have not been included. No exploration of CT with either H or He from the $2s\,2p\;^3P^o$ metastable term of \ion{N}{iv} is discussed in those works, nor were any other calculations found. \ion{N}{v} CT with H was included from \cite{stancil1997n5}.

\subsubsection{Neon}

The ionisation potential of \ion{Ne}{i} is reasonably close to He and so CT with He is the only important charge transfer process to consider. The excited levels in \ion{Ne}{i} and \ion{Ne}{ii} are so far from the ground that CT is not relevant for those levels in the solar setting. Radiative CT ionisation dominates at low temperatures, and the rate coefficients from \cite{liu2010ne1rad} for the ground level were incorporated. For collisional CT ionisation of \ion{Ne}{i}, the lowest energies given by \cite{liu2010ne1col} for the cross sections are above the region where the collisional process begins to take over from the radiative process. Consequently, to provide a smooth transition between the radiative and collisional regimes, the cross sections for collisional CT were supplemented with those at low energies from \cite{zygelman1986}. For CT with H from \ion{Ne}{iii} the only theoretical study is by \cite{heil1983}. The cross section for the process is ten orders of magnitude lower than for other ions, and so this has not been included. This is confirmed by the experiment of \cite{mroczkowski2003}, which only begins at energies of 160\,eV, while \cite{forster1991} also estimate the rate coefficient would be many orders of magnitude below that of the electron collisional processes. \cite{zhao2006} provides radiative CT with He rate coefficients for the ground and metastable levels of \ion{Ne}{iii}, and \cite{imai2003} is used for collisional CT. In the latter work, the cross sections do not go to low enough energies for the rate coefficients required for the solar atmosphere. For the ground level, the cross section drops very quickly at the lowest energy they show, meaning CT should be irrelevant at low energies, and so no addition was made to the data they give. For the metastable levels, they show that theory is in close agreement with the Okuno \& Kaneko (1983, unpublished) experiment, for which they estimate that there is an equal population of ions in the two metastable terms. Moreover, since the theoretical calculation shows that the metastable levels have similar cross sections at low energies and the ground level would contribute nothing to the experimental cross sections, then the Okuno \& Kaneko values \citep[as reported by][]{imai2003} at energies below the theoretical ones are included in order to calculate the rate coefficients. Double electron CT rate coefficients from \cite{imai2003} were also included. \cite{rejoub2004} provides the cross sections from which rate coefficients were obtained for CT from \ion{Ne}{iv} with H.

\subsubsection{Silicon}

For silicon, rate coefficients for CT ionisation of the neutral atom were obtained from \cite{kimura1996si1} for both the ground and metastable terms reacting with hydrogen. CT reactions from the ground and metastable terms of \ion{Si}{ii} were included back to the ground in the neutral atom by using detailed balance. CT ionisation from the metastable term in \ion{Si}{i} ends on the short-lived, $3s\,3p^2\;^2D$ term, and so no reaction from \ion{Si}{ii} to that term needs to be included. The cross sections of \cite{bacchus1998si3} and \cite{clarke1998} are within 20\;per cent of each other for reactions involving the ground of \ion{Si}{iii}. The latter work has a wider energy range, however, and includes CT from the metastable term. The rate coefficients from the latter work, therefore, were incorporated. Notable differences were found between the cross sections of \cite{wang2006si4} and \cite{liu2014si4} for CT with H from \ion{Si}{iv}, but the former work covered an energy range more suitable for this work and so their data were used. For CT with He from this ion, data from \cite{stancil1999si4he} were used because they provide cross sections which make it possible to determine the reverse, CT ionisation rate coefficients from both the ground and metastable levels of \ion{Si}{iii}, which is important for this ion. For CT reactions with H from \ion{Si}{v} no data was found except for that in \cite{kingdon1996}, but this is only valid up to 50000\,K, and so will not be relevant in the solar case. \cite{stancil1997si5he} was used to incorporate rate coefficients for CT with He.

\subsubsection{Sulphur}

The literature on CT involving sulphur is much less extensive than for the other elements explored in this work. This particularly affects level resolved modelling because, in the main, the few existing works do not calculate data for metastable levels. The only exception to this is the work of \cite{zhao2005s1} for \ion{S}{i}. They have CT ionisation from the ground and $3s^2\,3p^4\;^1D$ metastable term to the ground and metastable terms of \ion{S}{ii}. This is sufficient to provide all the relevant reactions for these two charge states, and so CT reactions back into \ion{S}{i} for all of these transitions using detailed balance were included in the model. Rate coefficients for CT from the ground of \ion{S}{iii} were derived from the cross sections of \cite{christensen1981} and \cite{bacchus1993s3}, who covered different energy regimes. The composite cross section shows good agreement between the two works. \cite{bacchus1993s3} show that the reactions end on the $3s^2\,3p^3\;^2D^o,\;^2P^o$ metastable terms in \ion{S}{ii}, and give ratios of the cross sections for both reactions. To calculate rate coefficients for the reverse reactions from each of the metastable terms in \ion{S}{ii}, the final cross section for CT from the ground of \ion{S}{iii} was split by the average ratio of the cross sections into the two metastable terms. The reverse reaction rate coefficients were then calculated from these. No data was found for CT from the metastable levels of \ion{S}{iii}. CT with He is covered by \cite{zhao2005s3he}, again just from the ground term. Studies for \ion{S}{iv} by \cite{labuda2004s4} for H and \cite{bacchus1998s4he} for He cover the keV energy range, which is too high for solar studies. \cite{kingdon1996} has rate coefficients for H from \cite{butler1980}, but they are only valid up to 30000\,K. The CT rate coefficients with He from \cite{butler1980} only go up to 30000\,K, also, although the CT ionisation rates go up to 300000\,K. Consequently, the CT ionisation rate coefficients were used, and rate coefficients for CT were obtained by detailed balance. However, the detailed balance CT rates coefficients at low temperature calculated in this way are an order of magnitude lower than the CT rate coefficients explicitly calculated by \cite{butler1980}. Capture to other levels than the ground of \ion{S}{iii} may be possible from the ground of \ion{S}{iv}, which the CT ionisation rate coefficients will not include. Consequently, these rate coefficients present the minimum effect this process will have between these two ions. Finally, \cite{stancil2001s5} provide rate coefficients for \ion{S}{v} in reactions with H and \cite{wang2002s5he} for reactions with He. Both of the works are included, although there are no rates available for the metastable levels. It will be reported in the results, in Sect.\;\ref{sec:sresults}, to what extent the missing CT rates from the metastable levels affects the results.

\subsubsection{Magnesium}

No works dedicated to charge transfer with magnesium were found. The only available rate coefficients are in \cite{kingdon1996}, and the references from which they derive their data. The focus of the present work for magnesium is to have ion populations for the higher charge states, primarily because no observations of the low charge ions were found while preparing for the comparison with observations. Based on the results for the other elements, it is unlikely that the higher charge states are affected by photo-induced processes and charge transfer, and so these processes have not been included in the level-resolved CR model of magnesium. As a brief exercise, however, to assess how much the low charge states of this element may be affected by the processes included in the new modelling, the rate coefficients from \cite{kingdon1996} were incorporated into the coronal approximation. To complement this, the cross sections of \cite{verner1996} were used to calculate photo-ionisation rates.

\section{Results for ion formation}
\label{sec:ionresults}

In this section, the results of the collisional-radiative modelling will be presented for each element. All CR models were run at a constant pressure of 3$\times$10$^{14}$\,cm$^{-3}$\,K. This is the pressure in the model atmosphere of \cite{avrett2008}, from which the hydrogen abundance and fractional populations were taken. It accords with the values obtained by \cite{warren2005} from lines ratios in the quiet-Sun TR. The CR models which only include the processes given in Sect.~\ref{sec:lrmethods} are referred to as electron collisional models, and those which include all the processes described in Sect.\;\ref{sec:methods} are called full models in the rest of this work. These are compared with the results from the coronal approximation of \textsc{Chianti}, which only considers ions in the ground state, and EII and DR and RR as the processes determining the ionisation equilibrium. To aid in understanding the reasons for the effects seen in the models, a list of ionisation and recombination rates for each atomic process are given in Tabs\;\ref{tab:ionrates}\;and\;\ref{tab:recrates}, for the low charge states. The rates are given at temperatures just above the peak formation temperature of the charge state listed, which is approximately the point where ionisation begins to have an effect against recombination. 

\begin{table}
	\caption{Comparison of total ionisation rates (in s$^{-1}$) from the ground level of low charge states at the indicated temperature (in K).} 
	\centering	
		\begin{tabular}{p{0.3in}ccccc}
			\hline\hline \noalign{\smallskip}
			Ion & $log\;T_e$ & CT & PI & EII \\
			\noalign{\smallskip}\hline\noalign{\smallskip}
			
			\ion{N}{i} & 4.1 & $4.8\times10^{-3}$ & $2.7\times10^{-2}$ & $3.6\times10^{-4}$ \\
			\ion{N}{ii} & 4.5 & - & $6.5\times10^{-3}$ & $1.8\times10^{-3}$ \\
            \ion{N}{iii} & 4.9 & - & $1.1\times10^{-3}$ & $2.7\times10^{-2}$ \\
            \noalign{\smallskip}
			\ion{Ne}{i} & 4.2 & $3.4\times10^{-11}$ & $1.4\times10^{-2}$ & $6.6\times10^{-6}$ \\
			\ion{Ne}{ii} & 4.6 & $4.8\times10^{-10}$ & $4.7\times10^{-3}$ & $1.8\times10^{-4}$ \\
            \ion{Ne}{iii} & 5.0 & - & $1.4\times10^{-3}$ & $1.3\times10^{-2}$ \\
            \noalign{\smallskip}
			\ion{Mg}{i} & 3.8 & 3.2 & $3.7\times10^{-2}$ & $2.0\times10^{-3}$ \\
			\ion{Mg}{ii} & 4.2 & $1.3\times10^{-3}$ & $4.6\times10^{-4}$ & $2.8\times10^{-3}$ \\
            \ion{Mg}{iii} & 4.9 & - & $2.9\times10^{-4}$ & $9.0\times10^{-5}$ \\
            \ion{Mg}{iv} & 5.3 & - & $1.5\times10^{-4}$ & $1.1\times10^{-2}$ \\
            \noalign{\smallskip}
			\ion{Si}{i} & 3.8 & 45 & 1.6 & $5.6\times10^{-4}$ \\
			\ion{Si}{ii} & 4.0 & 1.8 & $9.4\times10^{-4}$ & $5.6\times10^{-6}$ \\
            \ion{Si}{iii} & 4.5 & $3.8\times10^{-2}$ & $4.8\times10^{-4}$ & $2.7\times10^{-4}$ \\
            \ion{Si}{iv} & 4.8 & $1.1\times10^{-6}$ & $2.0\times10^{-4}$ & $2.5\times10^{-3}$ \\
            \noalign{\smallskip}
			\ion{S}{i} & 3.9 & $3.8\times10^{-1}$ & $1.8\times10^{-1}$ & $3.3\times10^{-4}$ \\
			\ion{S}{ii} & 4.3 & $3.4\times10^{-7}$ & $1.5\times10^{-3}$ & $2.3\times10^{-4}$ \\
            \ion{S}{iii} & 4.8 & $1.1\times10^{-2}$ & $7.5\times10^{-4}$ & $1.2\times10^{-1}$ \\

            \noalign{\smallskip}\hline
		\end{tabular}
	\label{tab:ionrates}
\end{table}

\begin{table}
	\caption{Comparison of total recombination rates (in s$^{-1}$) into the indicated ion from the ground level of the next higher charge state, at the indicated temperature (in K).} 
	\centering	
		\begin{tabular}{p{0.3in}ccccc}
			\hline\hline \noalign{\smallskip}
			Ion & $log\;T_e$ & CT & RR & DR \\
			\noalign{\smallskip}\hline\noalign{\smallskip}
			
			\ion{N}{i} & 4.1 & $7.0\times10^{-4}$ & $7.8\times10^{-3}$ & $9.0\times10^{-3}$ \\
			\ion{N}{ii} & 4.5 & $1.2\times10^{-1}$ & $1.1\times10^{-2}$ & $2.6\times10^{-2}$ \\
            \ion{N}{iii} & 4.9 & $2.1\times10^{-3}$ & $5.5\times10^{-3}$ & $5.6\times10^{-2}$ \\
            \noalign{\smallskip}
			\ion{Ne}{i} & 4.2 & - & $2.5\times10^{-3}$ & $2.0\times10^{-4}$ \\
			\ion{Ne}{ii} & 4.6 & $5.5\times10^{-8}$ & $4.3\times10^{-3}$ & $2.6\times10^{-3}$ \\
            \ion{Ne}{iii} & 5.0 & $3.8\times10^{-4}$ & $3.0\times10^{-3}$ & $2.5\times10^{-2}$ \\
            \noalign{\smallskip}
			\ion{Mg}{i} & 3.8 & $1.5\times10^{-3}$ & $2.3\times10^{-2}$ & $5.9\times10^{-2}$ \\
			\ion{Mg}{ii} & 4.2 & $8.0\times10^{-5}$ & $1.2\times10^{-2}$ & $3.4\times10^{-13}$ \\
            \ion{Mg}{iii} & 4.9 & - & $2.9\times10^{-3}$ & $2.3\times10^{-3}$ \\
            \ion{Mg}{iv} & 5.3 & - & $1.6\times10^{-3}$ & $1.4\times10^{-2}$ \\
            \noalign{\smallskip}
			\ion{Si}{i} & 3.8 & - & $2.5\times10^{-2}$ & $1.6\times10^{-1}$ \\
			\ion{Si}{ii} & 4.0 & 39 & $4.2\times10^{-2}$ & $2.1\times10^{-1}$ \\
            \ion{Si}{iii} & 4.5 & $5.0\times10^{-1}$ & $1.4\times10^{-2}$ & $2.2\times10^{-1}$ \\
            \ion{Si}{iv} & 4.8 & $1.6\times10^{-2}$ & $7.5\times10^{-3}$ & $2.3\times10^{-7}$ \\
            \noalign{\smallskip}
			\ion{S}{i} & 3.9 & $1.8\times10^{-7}$ & $1.4\times10^{-2}$ & $4.3\times10^{-2}$ \\
			\ion{S}{ii} & 4.3 & $5.2\times10^{-4}$ & $1.7\times10^{-2}$ & $1.2\times10^{-2}$ \\
            \ion{S}{iii} & 4.8 & $4.5\times10^{-4}$ & $6.8\times10^{-3}$ & $2.0\times10^{-1}$ \\

            \noalign{\smallskip}\hline
		\end{tabular}
	\label{tab:recrates}
\end{table}

\subsection{Nitrogen}
\label{sec:nresults}

\begin{figure}
	\centering
	\includegraphics[width=8.4cm]{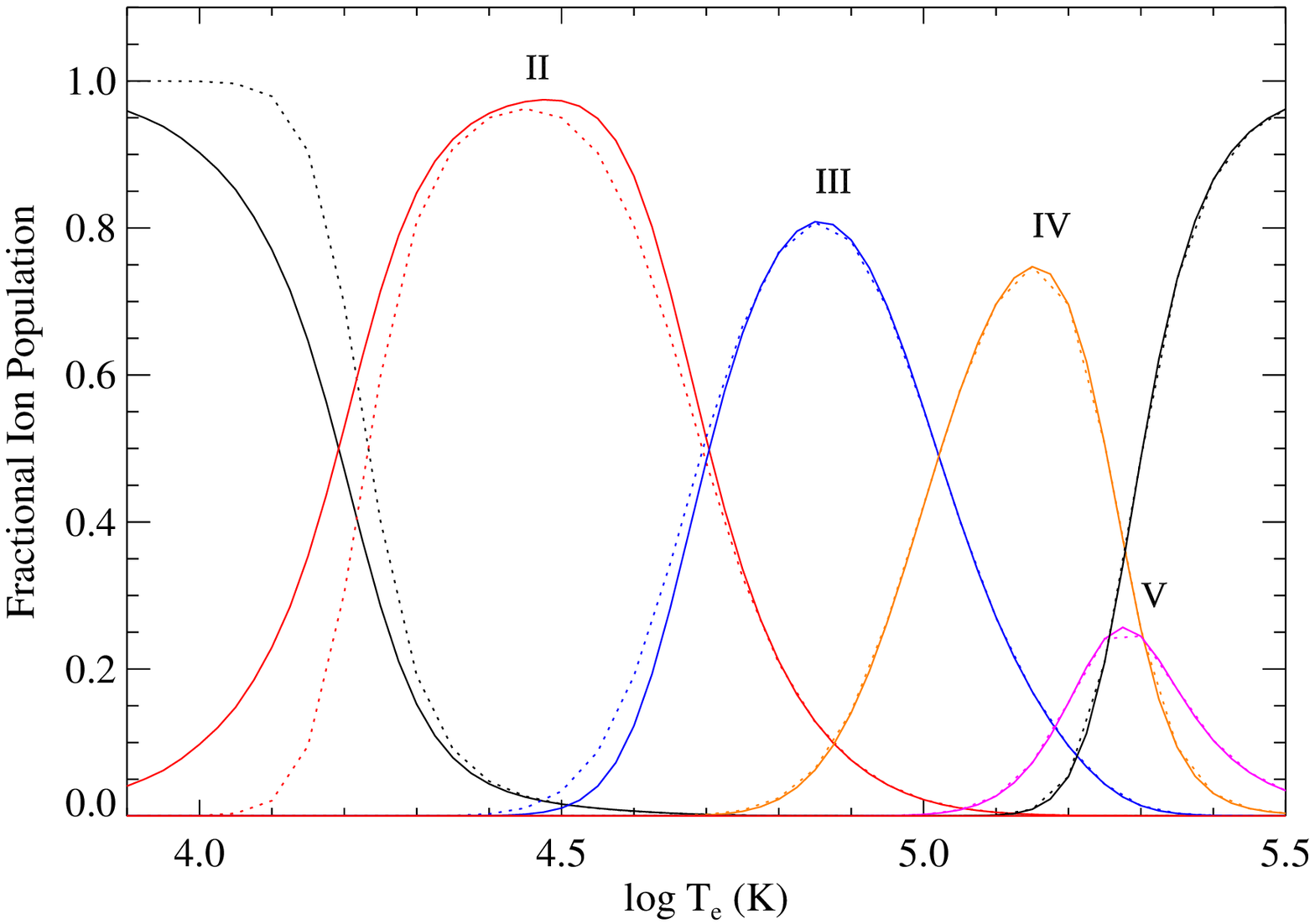}
	\caption[width=1.0\linewidth]{Coronal approximation for nitrogen: solid line - with CT included only, dotted - \textsc{Chianti} v.9.}
	\label{fig:ncrmct}
\end{figure}

\begin{figure}
	\centering
	\includegraphics[width=8.4cm]{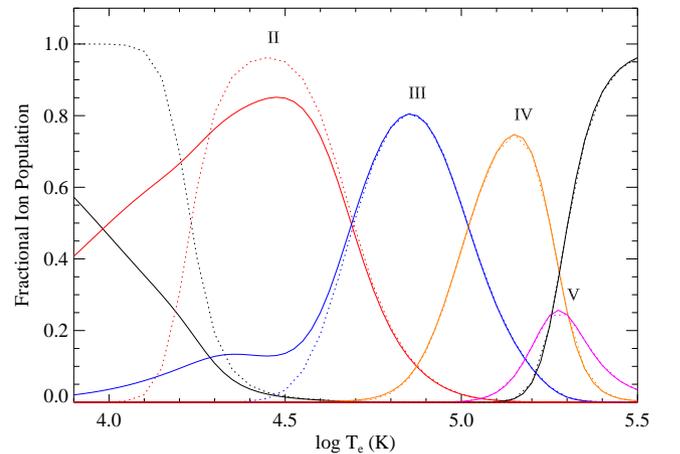}
	\caption[width=1.0\linewidth]{Coronal approximation for nitrogen: solid line - with PI included only, dotted - \textsc{Chianti} v.9.}
	\label{fig:ncrmpi}
\end{figure}

The way the ionisation equilibrium of nitrogen alters, with each of the processes added, mirrors closely the changes seen for oxygen in \cite{dufresne2020pico}. The difference mainly arises from charge transfer not being a resonant process for neutral nitrogen, (the threshold is 0.94\,eV). In contrast to \ion{O}{i}, then, CT ionisation reduces the population of \ion{N}{i} only by 10-20 per cent in the upper chromosphere. As with \ion{O}{ii}, the \ion{N}{ii} population is enhanced in the lowest parts of the solar transition region due to CT from \ion{N}{iii}. If CT is considered on its own, \ion{N}{ii} would have a near 100 per cent abundance at its peak temperature of 25000\,K and would form over a wider range than in models with electron collisional processes alone. This is shown in Fig.\;\ref{fig:ncrmct}, where CT from the ground level has been added to the coronal approximation modelling. The over-enhancement of \ion{N}{ii} through CT, at temperatures above the peak formation temperature, is stronger than the effect of CT on \ion{O}{ii}. The populations of \ion{N}{iii} are correspondingly depleted in this region.

When considering the effect of photo-ionisation on nitrogen alone, as shown in Fig.\;\ref{fig:ncrmpi}, it is possible to see the way nitrogen is affected in a similar fashion as oxygen. The PI threshold of \ion{N}{i} is 853\,\AA, and so it is ionised by much the same region as \ion{O}{i} (910\,\AA). Its cross section is 50 per cent higher than \ion{O}{i}, although its PI rate is only 20 per cent higher because it experiences less of the strong Lyman continuum that \ion{O}{i} experiences. PI confines the population of \ion{N}{i} towards the lower chromosphere, which means \ion{N}{ii} has a greater presence in the upper chromosphere compared to when electron collisional processes are considered alone. Because the continuum is lower near the threshold of \ion{N}{ii} (396\,\AA), and its threshold is further away from the \ion{He}{ii} 304\,\AA~line, \ion{N}{ii} is not so strongly photo-ionised as \ion{O}{ii}. Despite this, the 304\,\AA~line still accounts for 48 per cent of the PI rate. Considering PI alone, its peak abundance would be reduced to 85 per cent in the lower TR, which also enhances the population of \ion{N}{iii} in this region. PI reduces the peak population of \ion{N}{iii} by less than one per cent, and has no further effect in these quiet-Sun conditions on the higher charge states.

\begin{figure}
	\centering
	\includegraphics[width=8.4cm]{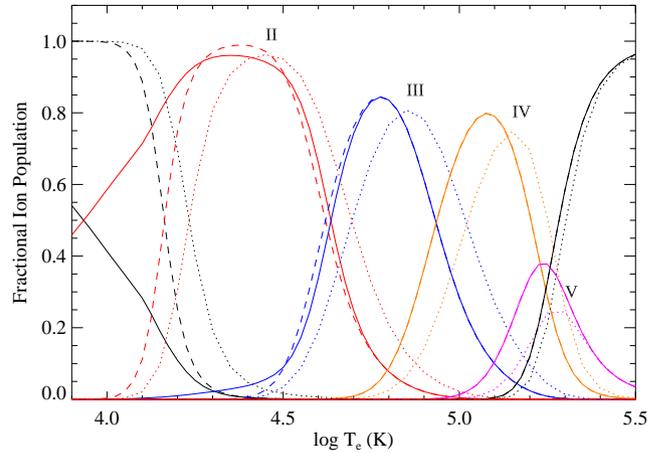}
	\caption[width=1.0\linewidth]{Ionisation equilibrium of nitrogen: solid line - full model, dashed - electron collisional model, dotted - \textsc{Chianti} v.9.}
	\label{fig:ncrmlrpict}
\end{figure}

When all of the atomic processes are combined in the full model, as illustrated in Fig.\;\ref{fig:ncrmlrpict}, it is PI which dominates the formation of neutral nitrogen compared to CT, which dominates in oxygen. In the case of singly ionised \ion{N}{ii}, it is similar to \ion{O}{ii}, in that its population is much stronger at lower temperature through PI. However, CT from \ion{N}{iii} not only enhances the population of \ion{N}{ii} above its peak formation temperature, it also ensures that the reduction in the population by PI at the peak is not as pronounced. Overall, \ion{N}{ii} now has a broad presence which dominates the upper chromosphere and lower TR. Doubly ionised \ion{N}{iii} is similar to \ion{O}{iii} in the way that CT offsets the enhanced populations that PI would produce, but it still has a small population which reaches down into the chromosphere. All of the transition region ions are shifted to lower temperature and higher peak population with metastable ionisation and recombination and suppression of DR added to the model. Li-like \ion{N}{v} shows a reasonable enhancement, in line with the change seen in \cite{dufresne2020} for Li-like \ion{O}{vi}, which forms at a similar temperature.

\subsection{Neon}
\label{sec:neresults}

\begin{figure}
	\centering
	\includegraphics[width=8.4cm]{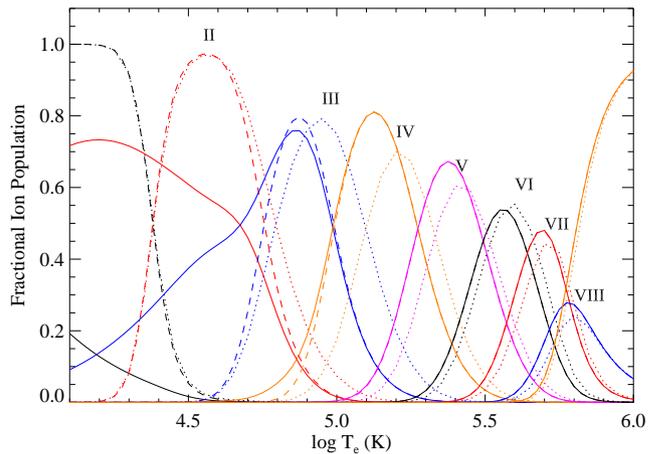}
	\caption[width=1.0\linewidth]{Ionisation equilibrium of neon: solid line - full model, dashed - electron collisional model, dotted - \textsc{Chianti} v.9.}
	\label{fig:necrmlrpict}
\end{figure}

The PI cross section of \ion{Ne}{i} rises substantially at energies further away from the threshold (21.6\,eV, 575\,\AA). The cross section peaks at the threshold of \ion{O}{ii}, and at this energy has a similar value to the \ion{O}{ii} cross section. Consequently, the PI rate for \ion{Ne}{i} is very similar to that of \ion{O}{ii}, and photo-ionisation by the 304\,\AA~line contributes almost 40 per cent to it. PI will then inevitably dominate \ion{Ne}{i} because the EII rate is low at the temperatures where it normally forms, as shown in Tab\;\ref{tab:ionrates}. \ion{Ne}{i} is correspondingly confined to the lower chromosphere and has a minimal population in the upper chromosphere, in comparison to the electron collisional model. The values of the \ion{Ne}{ii} PI cross section both at threshold and at its peak are similar to those of \ion{Ne}{i}, but, because its threshold (40.96\,eV, 302.7\,\AA) is just below the \ion{He}{ii} 304\,\AA~line, the PI rate is less than a half that of \ion{Ne}{i}. The reasonably strong continuum in the region 170-240\,\AA~does produce a strong rate for this ion, and it is still significantly stronger than EII. Figure\;\ref{fig:necrmlrpict} shows that \ion{Ne}{ii} is clearly affected by PI. A significant population of \ion{Ne}{iii} ensues from this, and is present not only at the base of the transition region, but down into the chromosphere. 

Tables\;\ref{tab:ionrates}\;and\;\ref{tab:recrates} show that CT has very little influence on the charge state distribution of Ne. This is principally because the dominant process for most ions is CT with He, but such CT usually only becomes significant at much higher temperatures. Indeed, the only influence CT has on this element is for it to reduce the small population of \ion{Ne}{iv} which would otherwise be present at lower temperatures from photo-ionisation of \ion{Ne}{iii}; it is CT with H which produces this, not CT with He. Overall, then, \ion{Ne}{i} becomes primarily confined to the lower chromosphere, while \ion{Ne}{ii} becomes an upper chromospheric ion. \ion{Ne}{iii} is now the primary ion present in the lower transition region. Regarding the higher charge states, the peak populations of \ion{Ne}{iv} and \ion{Ne}{v} are notably higher from including metastable ionisation and recombination in the modelling, as well as suppression of DR due to electron collisions. The population of \ion{Ne}{vi} is slightly reduced, while Li-like \ion{Ne}{viii} shows a small increase in population as a result of adding these processes to the modelling.

\subsection{Silicon}
\label{sec:siresults}

On its own, photo-ionisation would almost fully ionise \ion{Si}{i}, and \ion{Si}{ii} would have a fractional population of more than 80 per cent over the solar chromosphere. However, when the effect of charge transfer is taken into account, PI has no impact on the final charge state distribution. Table\;\ref{tab:ionrates} shows that CT ionisation is the dominant ionisation process for Si in the solar atmosphere up to \ion{Si}{iv}, confirming the findings of \cite{baliunas1980}. \ion{Si}{i} and \ion{Si}{ii} are affected by charge transfer ionisation with hydrogen. Because the threshold for the CT reaction from \ion{Si}{ii} back into \ion{Si}{i} is 5.4\,eV the rates are negligible and ionisation from \ion{Si}{i} is unopposed. In the very lowest regions of the atmosphere, the fractional abundance of ionised hydrogen, which is required for CT ionisation, is $10^{-4}$ or less. This, then, illustrates the strength of the CT ionisation rate coefficients compared to those for other neutral atoms and for other processes. Regarding the CT ionisation rate coefficients used in this work, \cite{cummings2002} reduced the same rate coefficients of \cite{kimura1996si1} by a factor of 2.9 for this ion because they believe the values may be overestimated due to the methods used by \cite{kimura1996si1}. They base this on the \cite{kimura1997} rate coefficients for neutral oxygen being this factor greater than those of \cite{stancil1999o1}. Even with the same reduction applied in this work, it would change little the almost complete depletion of \ion{Si}{i} over much of the chromosphere.

\begin{figure}
	\centering
	\includegraphics[width=8.4cm]{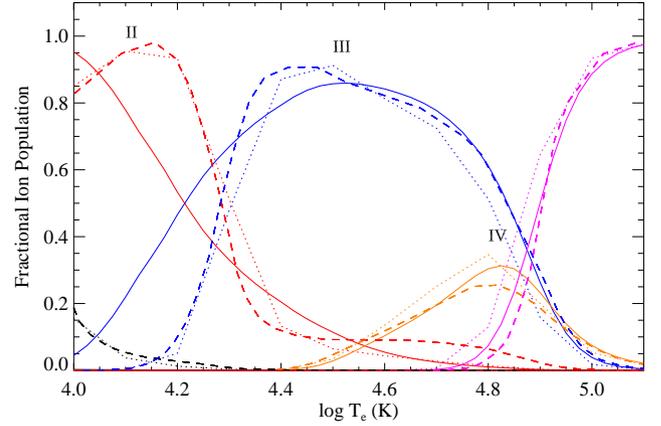}
	\caption[width=1.0\linewidth]{Coronal approximation for silicon: solid line - present work including charge transfer, dashed - Baliunas \& Butler, dotted - Arnaud \& Rothenflug.}
	\label{fig:sicrmct}
\end{figure}

The rates for CT ionisation out of \ion{Si}{ii} and CT back into it are four orders of magnitude greater than the other processes included here. The ionisation potential of \ion{Si}{ii} is 2.8\,eV greater than hydrogen, and so the stronger rate coefficients are for CT to take place from \ion{Si}{iii} into \ion{Si}{ii}, rather than the reverse reaction. However, in the increasing temperature in this part of the atmosphere where \ion{Si}{ii} and \ion{Si}{iii} form, the population of \ion{H}{i} gradually diminishes. Thus, CT ionisation out of \ion{Si}{ii} slowly takes over CT from \ion{Si}{iii} as temperature increases. Consequently, \ion{Si}{ii} forms over a broad temperature range. For \ion{Si}{iii} the ionisation potential (33.5\,eV) is too great for CT ionisation with H to occur; the rate is solely comprised of CT ionisation with He. This has a substantial impact on the formation of \ion{Si}{iv}, enhancing its population both at the peak in ion formation and at lower temperatures. In the ionisation region of \ion{Si}{iv} electron impact ionisation is the dominant process, and the usual ionisation equilibrium curves in electron collisional models are seen for \ion{Si}{v} and higher charge states.

Figure~\ref{fig:sicrmct} illustrates the ionisation equilibrium when the CT rates for the ground levels used in this work are added to the coronal approximation of \textsc{Chianti}. They are compared to the charge state distributions derived by \cite{arnaud1985} and \cite{baliunas1980}, both of which include CT processes. Despite the inevitable differences which will arise as a result of the different CT rate coefficients and ion fractions of H and He being used after such a long period has elapsed, many broad similarities in the curves are obvious, particularly for \ion{Si}{iii} and \ion{Si}{iv}. \cite{arnaud1985} use a virtually flat rate coefficient of $10^{-11}$\,cm$^3$\,s$^{-1}$, as derived by \cite{baliunas1980} at $10^4$\,K, for CT ionisation from \ion{Si}{i}, which explains the greatest difference in the curves. \cite{baliunas1980} state that if the rate coefficient for \ion{Si}{i} was greater than their estimate the ion would be significantly depleted.

\begin{figure}
	\centering
	\includegraphics[width=8.4cm]{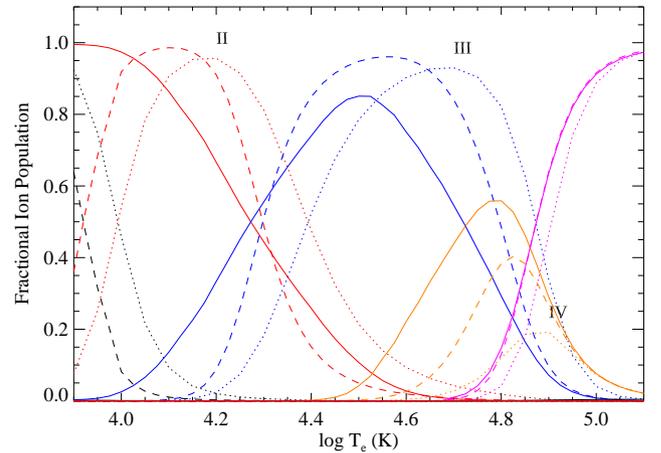}
	\caption[width=1.0\linewidth]{Ionisation equilibrium of silicon: solid line - full model, dashed - electron collisional model, dotted - \textsc{Chianti} v.9.}
	\label{fig:sicrmlrpict}
\end{figure}

There is a subtle, but noticeable, change in the ion balance when ionisation and recombination from metastable levels are added. The populations of the metastable $^4P$ levels in \ion{Si}{ii} are 5 per cent and those of the $^3P$ levels in \ion{Si}{iii} are 8 per cent, at their respective peak formation temperatures. CT from the $^1S$ level and $^3P$ levels in \ion{Si}{iii} ends on the $3s\,3p^2\;^2P$ and $^2D$ levels in \ion{Si}{ii}. In that case, there is no CT ionisation from the $3s\,3p^2\;^4P$ levels in \ion{Si}{ii} once they become populated. This reduces the effective ionisation rates out of \ion{Si}{ii} in the level resolved picture, compared to the coronal approximation. Furthermore, the CT rate from the $3s~3p~^3P$ levels in \ion{Si}{iii} are two orders of magnitude lower than that from the ground level. As the metastable levels become populated, it reduces the effective recombination rate into \ion{Si}{ii}. Given the influence of the charge transfer process on Si, then, metastable ionisation and recombination shifts \ion{Si}{ii} and \ion{Si}{iii} to slightly higher temperatures, when compared to the coronal approximation. This contrasts with the normal effect which metastable levels have, that is, to shift ion formation to lower temperature. In the formation of \ion{Si}{iv}, although the ionisation rate of the $^3P$ levels in \ion{Si}{iii} are more than two orders of magnitude lower than that of the ground, EII starts becoming more important at these temperatures. The greater effective ionisation rate out of \ion{Si}{iii} leads to a small enhancement in the population of \ion{Si}{iv} compared to the coronal approximation.

The other important process which substantially affects the charge state distribution is suppression of dielectronic recombination. As with other elements, this affects the higher temperature ions more. \ion{Si}{ii} is barely affected by this process because the DR rates into it are inconsequential compared to the CT rates. The formation of \ion{Si}{v} is also not affected because the DR rates into \ion{Si}{iv} are so low in this region. The DR rate from \ion{Si}{iv} is comparable to the CT rate, and suppressing this rate has a substantial effect on the formation of \ion{Si}{iii} and \ion{Si}{iv}. Much like the effect this process has on \ion{C}{iv}, the fractional population of \ion{Si}{iv} at its peak rises from 0.32 to 0.57. Since the process does not alter \ion{Si}{ii}, the overall effect on \ion{Si}{iii} is to substantially narrow the temperature range over which it forms. The fractional abundance of \ion{Si}{iv} seen here is similar to the ionisation equilibrium of \cite{kerr2019}, who adapt the coronal approximation from \textsc{Chianti} by including the charge transfer rate coefficients of \cite{arnaud1985} and DR suppression factors from \cite{nikolic2018}.

The final ionisation equilibrium for Si with all the effects modelled in this work is shown in Fig.\;\ref{fig:sicrmlrpict}. The changes compared to the coronal approximation of \textsc{Chianti}, which includes only collisions involving electrons, are striking. \ion{Si}{i} has much less presence in the solar atmosphere; these models suggest it is entirely confined low down in the chromosphere. \ion{Si}{ii} is shifted to much lower temperatures and now dominates the chromosphere, being almost entirely confined within it. \ion{Si}{iii} forms much lower in the TR, has a lower population at its peak, and its range of formation has been narrowed. Conversely, not only has the peak population of \ion{Si}{iv} almost tripled, the range of temperature over which it has significant populations has doubled; it now has a presence over most of the TR. As for the rest of the charge states of Si, the effects of density on electron collisional processes in the upper transition region are negligible; very little change is seen even in the populations of Li-like \ion{Si}{xii}.

\subsection{Sulphur}
\label{sec:sresults}

Given that the ionisation potentials of the lowest charge states of sulphur are similar to carbon, the effect of photo-ionisation on the ion balances for both elements is similar. Neutral sulphur has a larger PI cross section and there is a stronger continuum near its threshold, which mean that the PI rate is three times that of neutral carbon. This reduces the fractional population of neutral sulphur to less than 20 per cent in the upper chromosphere. Although the cross section of \ion{S}{ii} is almost the same as \ion{C}{ii}, the continuum at its threshold (529\,\AA) is not as strong as that experienced by \ion{C}{ii} at its threshold (506\,\AA). The PI rate for \ion{S}{ii} is half that of \ion{C}{ii}, and so, at the peak in ion formation, the fractional population of \ion{S}{ii} is only reduced by 5 per cent from PI. This is slightly less than the effect PI has on \ion{C}{ii}. The effect on \ion{S}{iii} is that it has a small presence in the upper chromosphere. Like the majority of doubly-charged ions explored in this series of works, there is no noticeable PI out of \ion{S}{iii} and higher charge states.

\begin{figure}
	\centering
	\includegraphics[width=8.4cm]{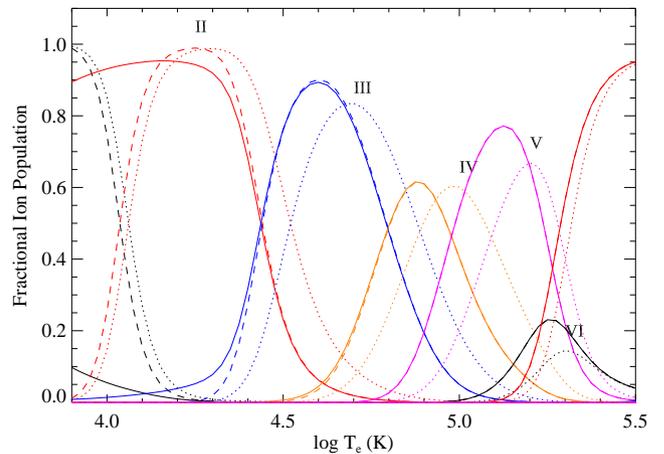}
	\caption[width=1.0\linewidth]{Ionisation equilibrium of sulphur: solid line - full model, dashed - electron collisional model, dotted - coronal approximation from present work.}
	\label{fig:scrmlrpict}
\end{figure}

The threshold for charge transfer ionisation out of the ground of \ion{S}{i} into the $^2P$ metastable term of \ion{S}{ii} is 0.2\,eV. For CT back into \ion{S}{i} from the ground of \ion{S}{ii}, the threshold is 3.2\,eV. Given that hydrogen is mostly ionised in the upper chromosphere, it is CT ionisation which is the main process governing the fractional abundance of \ion{S}{i}. Furthermore, it dominates all of the processes in the model when they are included together, such that \ion{S}{i} has a population of less than 10 per cent over the upper chromosphere. Surveying the CT rates for the other low charge states of sulphur in Tables\;\ref{tab:ionrates}\;and\;\ref{tab:recrates} shows that CT has no further influence on the ion balance, except for the small change produced by CT ionisation of \ion{S}{iii}. CT ionisation out of both \ion{S}{ii} and \ion{S}{iii} is due to collisions with He. Inevitably, the rate coefficients for this process will only be significant at higher temperatures. In addition, the presence of \ion{He}{ii} required for electron exchange means that CT ionisation will be a stronger reaction at the formation temperature of \ion{S}{iii}, rather than that for \ion{S}{ii}. \ion{S}{iii} is reduced in population by a few per cent from this, and the effect of the reverse, CT reaction is negligible in comparison. Although it was estimated that the CT rate coefficients used in this model for \ion{S}{iv} in collisions with He could be an order of magnitude lower than actual, the process is so weak compared to DR that the small depletion of \ion{S}{iii} by CT ionisation will be unaffected if higher CT rates from \ion{S}{iv} are included.

Regarding the discussion in Sect.\;\ref{sec:ctmethods} that few calculations for CT from metastable levels have been made and whether this would affect the model, \ion{S}{i} and \ion{S}{ii} are affected most by CT and \cite{zhao2005s1} did include reactions from metastable levels. These ions will, then, be modelled correctly. All the other reactions are too weak to influence the ion balance. Even with \ion{S}{iii}, the thresholds for CT ionisation from the $^1D$ and $^1S$ metastable levels to the available states in \ion{S}{iv} are 10\,eV higher than the threshold from the ground. Therefore, the rates from those levels are likely to be an order of magnitude lower than the ground rate, and so EII should be far stronger from these levels; the model should be unaltered even if metastable CT rates were included for this ion.

The new DR rates from \ion{S}{ii} incorporated in this work have a significant contribution at low temperature due to strong resonances close to threshold. The rate is three times greater than the RR rate over most of the formation temperature of \ion{S}{i}, causing \ion{S}{ii} to form at higher temperatures than indicated in the \textsc{Chianti} model. Since this low temperature component of DR is less suppressed at high densities, it means that the only shift in \ion{S}{ii} to lower temperature in the electron collisional model arises from the change in effective ionisation and recombination rates owing to the presence of metastable levels. Furthermore, ionisation out of \ion{S}{ii} is increased as metastable levels become populated, and the range over which \ion{S}{ii} forms is narrowed. The effect of density on electron collisional processes also alters the formation of \ion{S}{iii-vi}. The change to \ion{S}{iii} is mostly caused by the presence of metastable levels; \ion{S}{iv} is affected equally by this and suppression of DR, while \ion{S}{v} and \ion{S}{vi} are primarily influenced by DR suppression. The charge states above \ion{S}{vii} are little affected by these processes when compared to the coronal approximation.

\subsection{Magnesium}
\label{sec:mgresults}

\begin{figure}
	\centering
	\includegraphics[width=8.4cm]{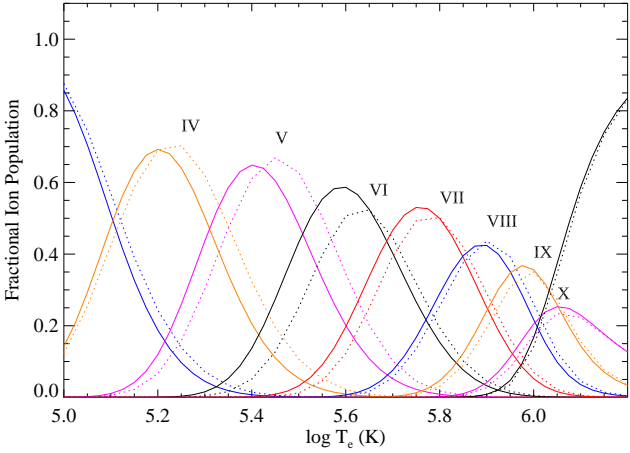}
	\caption[width=1.0\linewidth]{Ionisation equilibrium of the higher charge states of magnesium: solid line - electron collisional model, dotted - \textsc{Chianti} v.9.}
	\label{fig:mgcrmlrdr}
\end{figure}

There are a number of observations for lines emitted by the higher charge states of Mg in the solar transition region, and so the main reason for modelling the element is to assess if the emission is altered in the new modelling. In Fig.\;\ref{fig:mgcrmlrdr} it can be seen that \ion{Mg}{iv-vii}  are somewhat altered by the inclusion of density dependent processes, in that they are shifted slightly to lower temperature and their peak abundances have changed to a small degree. Although \ion{Mg}{viii-xi} are less affected, the changes to Mg are greater than those which occur for the higher charge states of Si and S.

\begin{figure}
	\centering
	\includegraphics[width=8.4cm]{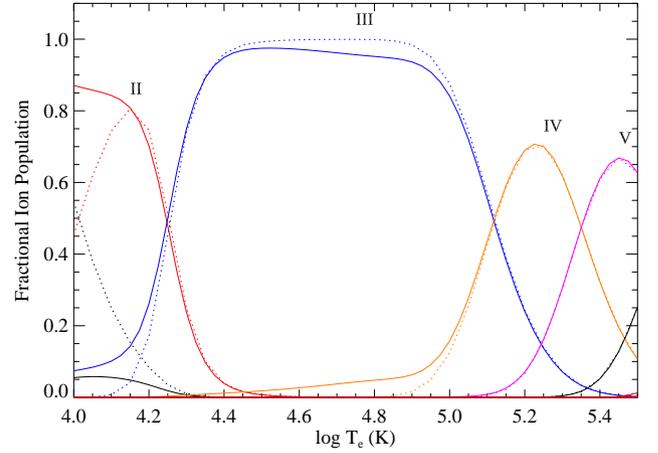}
	\caption[width=1.0\linewidth]{Coronal approximation of the low charge states of magnesium: solid line - with PI and CT included, dotted - \textsc{Chianti} v.9.}
	\label{fig:mgcrmpict}
\end{figure}

As a brief exercise, to see whether the low charge states of Mg may be affected by new atomic processes, PI and CT for the ground levels were included in the coronal approximation of \textsc{Chianti}. Of all the CT rate coefficients for Mg compiled by \cite{kingdon1996}, the only ones which cover ion formation at solar temperatures are those for \ion{Mg}{i} and \ion{Mg}{ii}. It is seen in Figure\;\ref{fig:mgcrmpict} that \ion{Mg}{i} is almost entirely depleted by charge transfer ionisation, such that \ion{Mg}{ii} is present at much lower temperatures. CT ionisation and, to a lesser degree, PI from \ion{Mg}{ii} contribute to EII, producing a greater population of \ion{Mg}{iii} at lower temperature. It is noted that Ne-like \ion{Mg}{iii} is affected by PI because the cross section is two orders of magnitude greater than those of the two lower charge states. This reduces its population in the TR by nearly ten per cent, and \ion{Mg}{iv} has a population down to the lowest parts of the TR.

\section{Level populations and line contribution functions}
\label{sec:levelresults}

As noted in \cite{dufresne2020pico}, the radiative decay rates between the ground and metastable levels are so low that the corresponding photo-excitation (PE) rates between those levels are not sufficient to alter their populations, and so the process does not affect the ion balances. However, it can still affect line emission in the atmosphere by altering level populations within ions. This section describes what effect PE has on individual ions, and also gives an indication of how line emission may be affected by all the processes modelled in this work by looking at selected line contribution functions.

\subsection{Nitrogen}

Nitrogen is less affected by photo-excitation than carbon because its ionisation potential is higher and the flux is not as strong at the shorter wavelengths required to excite its levels. All the terms in the $2s^2\,2p^2\,3s$ configuration are some of the most enhanced, with enhancement in the level populations being 10-20 times stronger than without PE. The $^4P$ term in this configuration is responsible for emitting the lines around 1200\,\AA, and is obviously enhanced by the continuum around the \ion{H}{i} Lyman-$\alpha$ line. Similarly, the $^2D$ term is connected to the $2s^2\,2p^3\;^2D^o$ metastable term by a transition of 1243\,\AA, where the continuum has the same strength. The $^2P$ term in the configuration is responsible for emitting the lines near 1160\,\AA. It is photo-excited from the $2s^2\,2p^3\;^2D^o$ metastable term by a transition of 1492\,\AA, at which wavelength the continuum is as strong as that around the Lyman-$\alpha$ line. The $2s\,2p^4$ and $2s^2\,2p^2\,3p$ configurations which lie between the terms of the $2s^2\,2p^2\,3s$ configuration are enhanced by 50 per cent up to a factor of three. This includes the $2s\,2p^4\;^4P$ term which emits at 1134\,\AA; it is enhanced by 50 per cent. All the higher levels, which arise from the $2s^2\,2p^2\,4s,\,3d$ configurations, are enhanced by factors of 3-40.

\begin{figure}
	\centering
	\includegraphics[width=8.4cm]{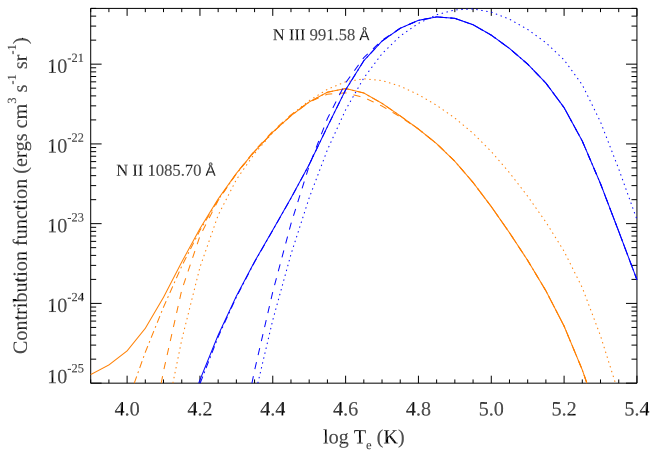}
	\caption[width=1.0\linewidth]{Contribution functions of \ion{N}{ii} 1085.70\,\AA~(orange) and \ion{N}{iii} 991.58\,\AA~(blue) lines: solid line - full model including PE, dash-dotted - full model without PE, dashed - electron collisional model of Dufresne et al., and dotted - \textsc{Chianti} v.9.}
	\label{fig:ncontribs}
\end{figure}

At the peak in ion formation of \ion{N}{ii}, the populations of all its levels are altered by 10 per cent at the most. Considering the contribution function of the 1085.70\,\AA~line (formed from the $2s^2\,2p^2\;^3P_2\,-\,2s\,2p^3\;^3D^o_3$ transition), for example, Fig.\;\ref{fig:ncontribs} indicates that  emission of the UV lines may be altered to a only small degree by both PI and PE despite the ion being formed at much lower temperature. The effect of charge transfer can be seen in the contribution function of this line, in that the peak in the contribution function coincides with the temperature at which \ion{N}{ii} populations are enhanced by charge transfer from \ion{N}{iii}. The results of the new modelling for \ion{N}{iii}, as shown in Sect.\;\ref{sec:nresults}, are clearly reflected in the contribution function for the $2s^2\,2p\;^2P^o_{3/2}\,-\,2s\,2p^2\;^2D_{5/2}$ 991.58\,\AA~line from this ion. It is reduced by charge transfer just below the peak, but emission obviously extends down to lower temperature through PI.

\subsection{Silicon}

As mentioned in Sect.\;\ref{sec:lrmethods}, level populations of neutral silicon will not be considered in detail because of the EIE data used, but, because ratios of level populations can be used to assess the effect of photo-excitation, a few remarks may be made on how it impacts this ion. Its levels are enhanced by factors of 10-100 at 10000\,K, while the terms of the $3s^2\,3p\,4s$ configuration, which are the first terms above the metastable, are enhanced by up to 1000 times. The higher terms of this configuration are responsible for lines around 2500\,\AA~and 3000\,\AA. Much of the photo-excitation comes from the ground, but the metastable $3s^2\,3p^2\;^1D,\;^1S$ terms do contribute to this process, as well. The overall effect is a considerable level of enhancement, but whether it is sufficient to offset the depletion of the ion by CT, as shown in Sect.\;\ref{sec:siresults}, would need to be tested.

The ionisation potentials of the third row elements are lower than those in the equivalent sequence in the second row. The low charge states of these elements should be affected by PE more, given that the solar flux is much stronger at the longer wavelengths required to photo-excite these ions. Considering the level populations of \ion{Si}{ii} at 13000\,K, (the temperature normally associated with the formation of its UV lines in the coronal approximation,) almost all levels populations are raised by 50\;per\;cent up to a factor of three. Photo-excitation comes from the $3s^2\,3p\;^2P^o$ ground term in this ion. Because of the difference in populations of the levels in the ground, the enhancement to excited levels can be different within the same term, potentially affecting line emission. For instance, the $3s^2\,3d\;^2D_{3/2}$ level, which is responsible for emitting the 1260.42\,\AA~and 1265.00\,\AA~lines, is enhanced by a factor of two, while the upper $^2D_{5/2}$ level of the term, which emits at 1264.74\,\AA, is increased by nearly a factor of three. Similarly, the enhancement to the $3s\,3p^2\;^2P_{3/2}$ level, emitting at 1190.42\,\AA~and 1194.50\,\AA, is 1.5 times greater than that of the lower, $^2P_{1/2}$ level, emitting the 1193.29\,\AA~and 1197.40\,\AA~lines. As described in Sect\;\ref{sec:siresults}, the new processes being included in the modelling indicates that \ion{Si}{ii} forms over most of the chromosphere. As a result, emission from this ion could be occurring at lower depths if PE is populating the levels.

In doubly-charged silicon PE has a much weaker influence on line formation. At the top of the chromosphere, where it has a much stronger abundance through CT, the levels are increased by 30\;per\;cent at the most by PE, while at its peak formation temperature of 35000\,K, the levels populations show almost no enhancement.

\subsection{Sulphur}

\begin{figure}
	\centering
	\includegraphics[width=8.4cm]{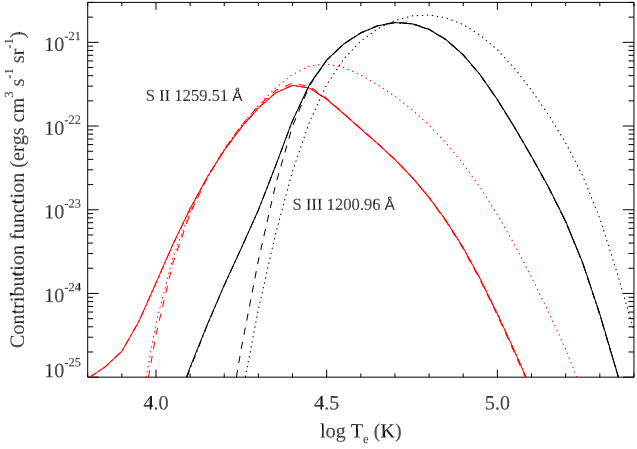}
	\caption[width=1.0\linewidth]{Contribution functions of \ion{S}{ii} 1259.51\,\AA~(red) and \ion{S}{iii} 1200.96\,\AA~(black) lines: solid line - full model including PE, dash-dotted - full model without PE, dashed - electron collisional model, and dotted - \textsc{Chianti} v.9.}
	\label{fig:scontribs}
\end{figure}

Neutral sulphur is represented in \textsc{Chianti} by only the ground and metastable terms, and so photo-excitation cannot be tested for any effects. Given that neutral sulphur does have some presence even in the upper chromosphere and that substantial enhancement is seen for other neutral species, it is reasonable to expect the level populations of \ion{S}{i} to be similarly altered. At the peak formation temperature of \ion{S}{ii} none of its levels which produce the routinely observed UV lines in the Sun show any enhancement. Figure\;\ref{fig:scontribs} shows that this is also true at lower temperatures at which the lines emit. Unlike other singly-charged ions that have greater populations from the current modelling lower in the atmosphere, the 1259.51\,\AA~line, arising from the $3s^2\,3p^3\;^4S^o_{3/2}\;-\;3s\,3p^4\;^4P_{5/2}$ transition in this ion, is barely enhanced by PE. The shift in formation of the ion to lower temperature, due to charge transfer and photo-ionisation, means there is some emission of this line lower in the atmosphere. The relatively slow decay rates between the first short-lived terms, $3s\,3p^4\;^4P,\;^2D$, and the ground and metastable terms are the reason for limited PE of this line. However, the $^4P,\;^4D,\;^4F$ terms arising from the $3s^2\,3p^2\,3d,\,4s$ configurations have strong radiative decay rates and their populations at the peak in formation temperature are all enhanced by up to a factor of three from PE.

In line with other doubly-charged ions, \ion{S}{iii} is unaffected by photo-excitation. As seen for the line at 1200.96\,\AA~from the $3s^2\,3p^2\;^3P_2\,-\,3s\,3p^3\;^3D^o_3$ transition, however, the presence of this ion lower in the atmosphere, through photo-ionisation of \ion{S}{ii}, allows electron collisions to populate the $^3D^o$ term at lower temperatures. This, along with the shift in ion formation to lower temperature caused by density effects on ionisation and recombination, may enhance its emission in the solar atmosphere because of the rising emission measure towards lower temperature.

\section{Conclusions}
\label{sec:concl}

Models which include all the main atomic processes that are likely to influence line formation have now been built for the main elements observed in the solar transition region. Through these, it is possible to confirm the claim of \cite{nussbaumer1975} that all transition region ions modelled here are affected when more advanced modelling is employed. Photo-ionisation on its own alters the charge state distributions of all the elements under consideration here. It can change the ion fractional populations of all low charge states, up to charge $+3$ for Ne and Mg and up to $+2$ for the other elements tested. It causes these charge states to have a presence at cooler temperatures than would otherwise be expected. 

Including charge transfer on its own, meanwhile, shows that all elements except neon are affected to some degree. In these cases, charge transfer ionisation depletes the neutral atoms. The third row elements are more susceptible to this, such that there is almost no presence of neutral atoms of these elements at solar temperatures for the conditions modelled in this work. Naturally, this establishes a significant presence of singly-charged ions at lower temperatures. For nitrogen, charge transfer can also produce an enhanced population of \ion{N}{ii} at higher temperatures than normal, just as it does with \ion{O}{ii}. 

With both PI and CT included together in the modelling, it generally occurs that these processes ionise the neutral species. For the singly-ionised species charge transfer from the charge state above tends to oppose photo-ionisation of the singly-charged ions. The only exception to these general trends is the formation of silicon. While photo-ionisation would ionise neutral silicon when considered on its own, it does not impact the charge state distribution when all the effects are included together. Not only does charge transfer entirely govern the formation of the silicon low charge states, but it is the only element for which the ion with charge $+3$ is entirely altered by the new effects.

When density effects on electron collisional processes are included, they play a part in altering the formation of both the low and high charge states which form in the transition region. Including metastable levels in the modelling forms a complex interplay of different rates into and out of each ion for the processes modelled. For ionisation by electron impact, rates for the metastable levels are enhanced relative to the ground. Charge transfer, however, is more state selective in comparison. A molecular state usually separates to final states which have similar total spin and orbital angular momentum as the initial states. Thus, the cross section for the process very much depends on similar terms being accessible in the neighbouring charge states within a reasonable threshold. This can make it a resonant process for one term, but a weak process for a neighbouring term. For instance, in \ion{O}{i} the charge transfer ionisation rates are substantially reduced for the metastable levels compared to the ground, whereas in \ion{Si}{i} they are much stronger. Photo-ionisation rates, although dependent on the ambient radiation field close to threshold, which may have strong lines present, do not show the same tendency to vary so dramatically from level to level as those of charge transfer.

It is primarily atoms which exhibit significant changes in their level populations through photo-excitation. Ionisation potential is the general guide which governs the strength of the effect. Singly-charged ions show only small changes due to photo-excitation, both at the peak in the ion populations and at lower temperatures, based on the selected contribution functions shown here. Doubly-charged ions are little affected by the mechanism.

Looking at all of the ionisation equilibria presented in this work, it is again clear that there is a dividing line at approximately 100\,000\,K, below which all of the effects included here alter the ion balances and should be incorporated into modelling. Above that temperature, the effect of higher density on collisions with free electrons is the only process affecting the ion balances compared to the coronal approximation. The changes shown here do depend on many factors, such as hydrogen abundance, opacity, pressure in the plasma, solar radiance, etc. Hydrogen abundance was taken from a hydrostatic model atmosphere, for which some of the parameters are determined semi-empirically. Density can also change between cell centres and granular networks, and does not exhibit an exactly constant pressure in some parts of the model atmosphere. The main aim here, however, has been to assess how much atomic modelling can change the predictions, and, once compared with observations, how effectively this modelling represents the broad conditions in the atmosphere. As noted in \cite{dufresne2020pico}, it is possible that the new processes included in these models are important for other regions in the solar atmosphere, such as active regions and flares, which have higher densities and radiances than the quiet Sun.

With the models built here, it will now be possible to establish a differential emission measure for the lower solar atmosphere which should better reflect the conditions present. From this, line intensities can be predicted and compared with observations. Not only will this determine the extent of the changes in the emission which arise from the modelling, but will also further go towards answering the issue raised at the beginning of this work, that is, where in the solar atmosphere the coronal approximation breaks down and more complex modelling should be employed. It should also be able to answer how deep into the atmosphere the modelling which includes all these processes in steady state equilibrium can reproduce emission in the quiet Sun. This assessment of the models will be the subject of a paper following this one.

\section{Data availability}

In cases where charge transfer and charge transfer ionisation rate coefficients were not given in the original sources, those calculated here from the published cross sections are made available. The ion fractional populations at constant pressure derived from the full models, in the \textsc{Chianti} `\texttt{.ioneq}' format, and DR suppression factors for \ion{S}{ii} are also made available. The data may be found at the CDS via anonymous ftp to cdsarc.u-strasbg.fr (130.79.128.5) or via http://cdsarc.u-strasbg.fr/viz-bin/qcat?J/MNRAS.

\section*{acknowledgements}
	
	Support by STFC (UK) via the Doctoral Training Programme Studentship is acknowledged, plus the support of a University of Cambridge Isaac Newton Studentship. GDZ acknowledges support from STFC (UK) via the consolidated grants to the atomic astrophysics group (AAG) at DAMTP, University of Cambridge (ST/P000665/1. and ST/T000481/1).
	
	Most of the atomic rates used in the present study were produced by the UK APAP network, funded by STFC via several grants to the University of Strathclyde. Acknowledgment is made of the use of the OPEN-ADAS database, maintained by the University of Strathclyde. \
	
	\textsc{Chianti} is a collaborative project involving George Mason University, the University of Michigan, the NASA Goddard Space Flight Center (USA) and the University of Cambridge (UK). \

\bibliographystyle{mnras}

\bibliography{pi_crm}

\bsp	
\label{lastpage}
\end{document}